\documentclass{article}

\usepackage{microtype}
\usepackage{graphicx}
\usepackage{subfigure}
\usepackage{booktabs} 

\usepackage{hyperref}

\usepackage[accepted]{icml2024}

\usepackage{amsmath}
\usepackage{amssymb}
\usepackage{mathtools}
\usepackage{amsthm}

\usepackage[capitalize,noabbrev]{cleveref}

\theoremstyle{plain}

\theoremstyle{definition}

\theoremstyle{remark}

\usepackage[textsize=tiny]{todonotes}

\usepackage{url}

\icmltitlerunning{The Reasonable Person Standard for AI}

\begin{document}

\twocolumn[
\icmltitle{The Reasonable Person Standard for AI}



\icmlsetsymbol{equal}{*}

\begin{icmlauthorlist}
\icmlauthor{Sunayana Rane}{princeton,chicago}
\end{icmlauthorlist}

\icmlaffiliation{princeton}{Department of Computer Science, Princeton University, Princeton, NJ, USA}
\icmlaffiliation{chicago}{University of Chicago Law School, Chicago, IL, USA}

\icmlcorrespondingauthor{Sunayana Rane}{srane@princeton.edu}

\icmlkeywords{Machine Learning, ICML}

\vskip 0.3in
]



\printAffiliationsAndNotice{}

\begin{abstract}
As AI systems are increasingly incorporated into domains where human behavior has set the norm, a challenge for AI governance and AI alignment research is to regulate their behavior in a way that is useful and constructive for society. One way to answer this question is to ask: how do we govern the human behavior that the models are emulating? To evaluate human behavior, the American legal system often uses the ``Reasonable Person Standard.'' The idea of ``reasonable'' behavior comes up in nearly every area of law. The legal system often judges the actions of parties with respect to \textit{what a reasonable person would have done under similar circumstances}. This paper argues that the reasonable person standard provides useful guidelines for the type of behavior we should develop, probe, and stress-test in models. It explains how reasonableness is defined and used in key areas of the law using illustrative cases, how the reasonable person standard could apply to AI behavior in each of these areas and contexts, and how our societal understanding of ``reasonable'' behavior provides useful technical goals for AI researchers. 

\end{abstract}

\section{Introduction}
AI governance and AI alignment research \cite{christian2020alignment} share the common goal of working towards safer, more reliable, and continually innovative AI. In recent years, breakthroughs in foundation models \cite{radford2021learning, ramesh2021zero,saharia2022photorealistic, ChatGPT} have surprised us by revealing capabilities hitherto thought of as only human. While these models have a long way to go, their increasing prevalence in human domains begs the question: how do we govern the human behavior they are emulating? 



One way to govern AI behavior (and define what kind of behavior we ought to optimize for as AI researchers) is to see what kind of behavior we expect of human counterparts, and how this behavior is regulated. Currently, AI models are understood to display behavior that is wildly out of the scope of normal human behavior \cite{raceai, uber}. But what if we could expect of a model, at the very least, exactly what we expect of reasonable human behavior in the same domain?

The notion of ``reasonable'' behavior seems fuzzy and as researchers in a quantitative domain, we may justifiably be initially skeptical that ``reasonable'' behavior can be defined, measured, and optimized for in a model. After all, human (and model) behavior is known to be unpredictable. However, we have a very good reason to believe that reasonable behavior can and should be a target for models -- the legal rules constraining human conduct in the United States are are built heavily on the assumption that we can recognize, and keep our own conduct within, the bounds of a collective societal understanding of reasonable behavior. This paper argues that the reasonable person standard should be an explicit goal of the AI research community, and an integral part of truly effective, practical AI safety measures.




The idea of a collective, societal notion of \textit{reasonable} behavior among humans has a long, storied foundation in every aspect of the American legal system. It can help us understand what society at large expects of an AGI's behavior, and how we can build those expectations into current models. Importantly, \textit{intelligent} behavior and \textit{reasonable} behavior are measured along two different axes in humans, and you can have people who exhibit one but not the other. Ideally, we want both. As most recent AI research has focused on creating increasingly \textit{intelligent} behavior, our metrics overly focus on this. Instead, it is arguably not just the \textit{number} of mistakes made but the \textit{type} of mistake that is important to study and shape. Was it a ``reasonable'' mistake? Would a human be able to understand, work around, and mitigate the damage of such a mistake? 
 



\section{The Reasonable Person Standard}

The legal system often judges the actions of parties in a case with respect to \textit{what a reasonable person would have done under similar circumstances} \cite{torts-textbook}. Baked into this idea is a notion of common sense and good faith in our interactions with one another. 

\subsection{Origins of the Reasonableness Standard}

\begin{quote}
``The life of the law has not been logic: it has been experience.''  

\hspace{3mm}- Oliver Wendell Holmes Jr., \textit{The Common Law}
\end{quote}

Justice Holmes (one of the most well-known and well-regarded jurists of the 20th century) made this reply to those jurists who believed that the law was a science of pure logic \cite{langdell}, and that once all the logical rules could be quantified legal decision-making would be straightforward. 


Instead of an exacting logic based on pre-existing experiences, Holmes argued, law was about adapting to new experiences and eventualities, and thus required the degree of flexibility needed to evolve to these new experiences and eventualities. Holmes and other legal realists argue that the \textit{common law}, the body of judge-made law built on individual cases and precedents, was intended to provide this flexibility. The American legal system incorporates substantial elements of English common law, which was used across the courts of English monarchs and brought to the American colonies at the time of their founding \cite{torts-textbook}. Many fundamental areas of law today still draw extensively from their common law roots: torts, contracts, and even criminal law all retain elements of common law precedent. 

The reasonable person standard developed first in common law \cite{garrett2017constitutional}, and therefore maintains influence in both European law  \citep{zorzetto2015italian}, and American law today. It was designed to be flexible, pertaining to the needs of every situation and the details of every case, while still being consistent with common law precedent (a case's decision with respect to reasonable behavior largely had to be consistent with past cases' rulings on reasonable behavior, so as to create a standard that people could understand and follow). 

\subsection{Inferences of what is reasonable}
Judges and juries (and all people who wish to follow the phrasing of the law in their everyday decisions so as to avoid breaking the law) are called on to determine what they believe to be reasonable conduct, and act within the boundaries of how a reasonable person could possibly behave. Few disputes ever lead to a lawsuit and even fewer are actually litigated in court, indicating that even though it seems like a fuzzy concept, most of the time our collective dependence on a mutual understanding of reasonable behavior does work.

 All of a judge's or jury's past experiences, social conditioning, curricular learning, etc. (training data, if you will)  are brought to bear every time they rule on whether someone's behavior is outside the bounds of how a reasonable person would behave, interpret a word, phrase a contract, take care not to hurt someone, etc. In machine learning terms, a crude analogy is conditioning through training data, but this is just a start (at least in current models). At the very least, the vast majority of people have an idea of what they construe as reasonable behavior. This understanding is demonstratable, explainable, and dependably communicable, with a safe assumption of some shared understanding of reasonableness. This is out of scope of what models currently do, although they are modeling their training data distribution. Models may well already have the building blocks (architecturally and in terms of data) to achieve a human-like understanding of reasonableness, but we won't know unless we specifically characterize, probe for, and and optimize for it.

\subsection{Rules vs Standards}
In the law, we often make a distinction between rules and standards. Rules are specific, inflexible, and apply regardless of the context of the individual circumstance (``driving in excess of 25 miles per hour is not allowed''). Standards are designed to be more flexible, because they are understandably context-dependent (``driving at excessively high speeds during inclement weather is not allowed''). Rules are often easier to implement, whereas standards are often more inclusive of all circumstances \citep{kaplow2013rules}. It has even been suggested by some legal scholars that the distinction between rules and standards will disappear in the future as algorithms can take into account all circumstances and determine guilt \citep{casey2016death}. The implication is that a different reasonableness threshold could apply to each driver based on each situation, as determined by an algorithm that can take into account all the details of the particular circumstances in a way that a legal system would not be able to do for each case. 

As machine learning researchers, we of course have many reasons to be skeptical of this projection, and wary of the dangers of allowing algorithms to make decisions regarding guilt. For now, however, the reasonable person standard is indeed a \textit{standard}, designed to be flexible and incorporate context. We don't have any pre-set rules regarding \textit{exactly} how a reasonable person behaves, because \textit{all} possible reasonable actions and beliefs in \textit{every} possible situation would be impossible to nail down ex ante. However, the justice system strives to hold most everyone to the \textit{same} bar or objective standard of what a reasonable person could possibly have done under similar circumstances. 

It is important to note here that the reasonable person is not the \textit{average} person by any means. The average person may occasionally text while driving; the reasonable person would not take such an unjustifiable risk \cite{torts-textbook}.

In the following sections we characterize and illustrate the pervasive presence of ``reasonableness'' in many important areas of the law, and what we can learn from these instantiations as we decide how to optimize for, probe, and stress-test reasonable behavior in our models. For each body of law, we walk through several key cases in which the question of ``reasonable'' human behavior  was considered by the courts, and in each case we discuss the implications, opportunities, and challenges for AI behavior. 

\section{Contract Law and Language}

How do two human parties make an agreement, and thus align using language to form a shared meaning that they mutually bind themselves to? Alignment achieved using language is enforced largely through contract law. Contracts are \textit{agreements} between human parties (or organizations). In creating one, it follows that two parties sufficiently aligned in intent and language. Contracts are often verbal, informal, and \textit{always} have implied terms that are based in the parties' (and the courts') understanding of what could reasonably have been meant by the words in the contract. 

As we discuss in this section, a simple phrase or sentence, verbally spoken, can constitute an enforceable contract even when parties later disagree as to the meanings of the words used in that sentence \cite{embry} -- what matters is whether a reasonable person could have interpreted the sentence in the way the aggrieved party interpreted it. If so, the reasonable person is rewarded and the contract is usually upheld.


For this reason, ``reasonableness'' has been a defining test in contract interpretation -- the court and the parties depend on a reasonable understanding of the ever-present implied, unwritten/unspoken terms of the contract, especially when the explicitly recorded terms are sparse.

Contracts were initially intended as a ``meeting of the minds'' \cite{farnsworth1967meaning}, a phrase which should be particularly salient to those of us who endeavor to align an artificial ``mind'' with a human one. It is interesting to consider what this would mean in the world of LLMs, when words are used skillfully but still with meanings that are arguably often very different from what we would expect of a reasonable human counterpart.


\subsection{Implied terms}

\begin{quote}
    ``The law has outgrown its primitive stage of formalism when the precise word was the sovereign talisman, and every slip was fatal. It takes a broader view to-day. A promise may be lacking, and yet the whole writing may be `instinct with an obligation,' imperfectly expressed.''

        \hspace{10mm}\parbox{\textwidth}{-Judge Cardozo \\ Wood v. Lucy, Lady Duff-Gordon, 1917}

\end{quote}

In the enduring case of \citet{1917wood}, the defendant Lady Duff-Gordon was sued for breaching an exclusive contract with an agent, Wood, in which he was to receive a cut of the profits from her sales of fashion merchandise. She went around this arrangement and withheld profits from ensuing sales. 

Lady Duff-Gordon's argument was that Wood technically didn't promise anything of value -- he didn't promise to market her merchandise, only to account for any payments received and file any protective patents necessary. She says that he never promised to make reasonable efforts to market her designs, and if he doesn't market them, there will be no payments to account for, in which case the remainder of his stated responsibilities become moot. Therefore, she claims, the promise was valueless and did not satisfy the formal criteria for a contract, even though they had agreed on the written words.

Judge Cardozo, however, in his quote above, rejected this logic. He held that there was an implied promise made to market her merchandise, even if Wood did not spell it out in so many words, and that this implied promise was of enough value to enforce the contract. In doing so, he solidified the idea of reasonable implications and assumptions arising from an agreement, outside of what is explicitly spelled out in the text.

In this case, Cardozo could infer what shared assumptions the parties held when entering the contract; he could do so because he could simulate both their understanding of the language in the contract, \textit{and} the understanding of the world that allows for shared meaning outside of the explicit text of the contract. In this case, the shared meaning was drawn from the most reasonable assumption of the intent of two parties who would enter into such an agreement. 

However, how can we even begin assigning reasonable shared meaning between humans and, say, LLMs or foundation models? The assumptions we make about Wood and Lady Duff-Gordon's motives and implied meanings in the text are not the kind of assumptions we can reasonably make about the language produced by an LLM. Human parties also frequently change their minds after the fact, lie, and obfuscate regarding their original intent when creating an agreement. However, we still find it at least tractable to resolve disputes about shared meaning between human parties, be cause we can make reasonable inferences about their hidden motives and beliefs; until we can do so with AI, we have not properly aligned AI with humans at the word- and concept-level. 

\subsection{Implied intent of words to a reasonable observer}

\begin{quote}
    ``Judicial opinion and elementary treatises abound in statements of the rule that to constitute a contract there must be a meeting of the minds of the parties, and both must agree to the same thing in the same sense. Generally speaking, this may be true; but it is not literally or universally true. That is to say, the inner intention of parties to a conversation subsequently alleged to create a contract cannot either make a contract of what transpired, or prevent one from arising, if the words used were sufficient to constitute a contract.''
    
    \hspace{9mm} - Judge Goode, Embry v. McKittrick, 1907
\end{quote}

The idea of implied shared meaning of words, as understood by a reasonable observer, prevails even when the words are vague and one party claims the agreement itself was never made. The question is not whether you intended to make a promise, but whether a reasonable person could have inferred a promise from your words.

This principle is illustrated in the case quoted above, \citet{embry}, where the plaintiff Embry met with the president of the defendant corporation and stated his intent to quit if he was not guaranteed a renewed contract for the next year. The president replied ``go ahead, you’re all right; get your men out and don’t let that worry you,'' from which Embry reasonably assumed that the contract had been renewed. He stayed, but was subsequently fired. The president later stated that he had meant to get back to the issue later, and never meant to enter into a renewed contract. However, the court ruled that the president's words were enough to induce a reasonable person to make a reasonable inference that the contract had indeed been renewed. Once again, the reasonable person standard prevailed, and the company was held to its promise. 

In another case which clarified the idea of implied intent to a reasonable observer, \citet{zehmer} tangled with the notion of a joking offer, and held that it was still enforceable if the recipient of the joking offer could reasonably believe that it was not in jest. The court said, ``We must look to the outward expression of a person as manifesting his intention rather than to his secret and unexpressed intention. `The law imputes to a person an intention corresponding to the reasonable meaning of his words and acts''' \cite{zehmer}, quoting \cite{firstnat}. This raises many problems for LLMs, which lie and hallucinate \cite{ji2023survey}, because it implies that regulation does not often deal with the tricky question of what is ``under the surface,'' including hidden motives and reasons for ``joking,'' lying, or even unintentionally misleading behavior. Instead, the law focuses on how those words and actions could have been reasonably interpreted, and does not punish a reasonable person for interpreting them as such. The liability is instead imposed on the party that, even unintentionally, led another to reasonably believe something; if it was reasonably believed, then they are bound to it. 

In a very recent testament to this principle, a Canadian tribunal held that an airline was liable for information its AI chatbot fabricated and provided to its customers \cite{airline-chatbot}. It did not matter whether the information was true or not, nor whether the chatbot was ``joking'' or hallucinating. As long as customers reasonably interpreted those words to be the truth, the airline was bound by its LLM's words.

Interpretability and explainability of AI models \cite{doshi2017towards} is an ongoing effort and we have a long way to go before we fully understand the causes for models' behaviors. In the meantime, when the general public are in contact with and affected by the words and actions of AI, the prevailing question will not be what the hidden \textit{intent} of the AI was (indeed it is difficult to define ``intent'' in terms of weights and biases), but rather how a reasonable person could have interpreted its words and actions. By incorporating the reasonable person standard into our research, we can better understand what behavior is reasonably expected of our models. Once we have begun to characterize and study AI behavior through this lens, it will be possible to build models that are more human-aligned.

\subsection{Reasonable interpretation of promises}
The reasonable person standard also works in the reverse -- the court will often decline to enforce a contract when one party claims there is a contract but the other claims there is not, and importantly, the court finds that no reasonable person could have interpreted the circumstances to mean that the claimed contract existed. In one of the more amusing cases regarding the reasonable person standard in contracts, \citet{pepsi}, a teenager sued Pepsi over a commercial that he felt implied that he could trade in a number of reward points for a Harrier Jet. 

The appeals court, affirming a prior district court ruling, concluded that ``no objective person could reasonably have concluded that the commercial actually offered consumers a Harrier Jet'' \cite{pepsi}. This case illustrates how we are also protected from unreasonable interpretations of our words, and how the surrounding context of the case matters tremendously in how the rules apply to it. While the reasonable person would have correctly interpreted the Pepsi commercial, it is not clear whether even our best models would appreciate the nuance and context sufficiently to fully understand it. We could imagine that many of our best models may interpret such a commercial literally. However, armed with this knowledge, if in the future we explicitly and methodically work towards AI models that meet the reasonable person standard, there is good reason to believe that we may be able to successfully apply our research community's substantial technical expertise to prevent precisely such unreasonable behavior.

\subsection{Multiple equally reasonable meanings}

The law also allows for cases where they may be multiple \textit{equally} reasonable interpreted meanings of agreed-upon language. In \citet{peerless}, a contract was made for cotton to be transmitted on a ship called Peerless. However, there were two ships with the same name, and the contract did not further stipulate which was meant. Later on, each party claimed he thought it was the other ship. The court, finding that the text was ambiguous enough to allow for two equally reasonable interpretations, held that the contract was not enforceable. This type of allowance should perhaps also be extended to ambiguity in LLM text; however, the bounds for this are yet to be defined, and our research community can help quantitatively determine what ``unambiguous'' word meanings should mean for an LLM interacting with a human (presumably this would change based on the context and the specific audience). Our research can also help determine how to preemptively address situations where multiple equally reasonable meanings can arise, and how to limit misunderstandings arising from such situations.






\section{Tort Law and Actions in the World}
Civil cases, which generally involve disputes between people or organizations, are often litigated under tort doctrines. Torts occur when one party is found to have harmed another by engaging in negligent conduct. The reasonable person standard is most often and explicitly referenced in tort cases. 
\begin{quote}
    ``Unless the actor is a child or an insane person, the standard of conduct to which he must conform to avoid being negligent is that of a reasonable man under like circumstances.'' 
    \\ 
    \hspace*{10mm} \citet{torts-restatement}
    \hspace*{10mm}  American Law Institute
\end{quote}
Tort law states that we all, by default, have a duty of reasonable care to one another -- in other words, we have a duty not to cause harm to one another. Whether or not a party is negligent comes down to whether they breached this duty of care by not taking the necessary and prudent precautions to prevent forseeable harm arising from their conduct \citep{wright2002justice}. The question courts ask: Would a reasonable person, under the same circumstances, have taken a precaution that you neglected to take that would have prevented these harms?

A pattern of erratic behavior does not excuse a human actor from being held to the reasonable person standard, and nor should it excuse an AI model and its human handlers. In the well-known case of \citet{neighbor-fire}, the defendant built a haystack on a property border, and his neighbor repeatedly told him it was a fire hazard. The defendant said he would chance it, and then built a chimney inside the haystack. The haystack subsequently caught fire, burning down his neighbor's property. The neighbor sued. The defendant claimed, in his defense, that he wasn't intelligent enough to have known that his actions would cause a fire and the damage that would be caused. The court ruled that this so-called lack of intelligence was not sufficient defense; a reasonable person should have known that these actions could lead to a dangerous fire, and would have taken precautions. The defendant was ordered to pay damages.

Our standards of reasonable behavior, and reasonable care, do not change when assessing parties who have a history of poor adherence to reasonable behavior; in other words, by and large, the reasonable person standard is an \textit{objective} one \citep{schwartz1989objective}. Applying this idea to AI behavior, we can see that the knowledge that AI models are known to produce erratic behavior is not a viable excuse; this knowledge does not then excuse the harm caused by future erratic behavior. 





\subsection{Foreseeability}
Key to tort liability is the idea of \textit{foreseeable} harm: we want to incentivize people to take steps to avoid harm they can foresee, but we do not want to punish people for harm that they could not possibly have foreseen. 

On first inspection, the legal construction of foreseeability seems like it could be used to absolve the humans behind AI models of responsibility for the model's behavior, simply because so much of AI behavior is \textit{not} forseeable or predictable (in fact, this is a key problem in AI safety -- when AI behavior does not conform to human expectations, the mismatch can cause dangerous outcomes) \citep{amodei2016concrete}. Machine learning models of all kinds have been shown to, at least occasionally, display such erratic behavior. 

However, the concept of foreseeability as constructed in tort law provides a way to deal with this type of behavior. A famous case now colloquially known as the ``flaming rat case,'' and more formally known as \citet{flaming-rat}, tested the extent to which the precise chain of events needs to be forseeable to impose liability. In this case, a company had instructed a young man to clean the company's coin-operated machines, which he was using gasoline to clean -- an inherently dangerous use of gasoline (especially in a room with a lighted gas heater) that made the situation accident-prone and thus constituted a breach of reasonable care by the company. 

What happened next, however, was rather unexpected. There was apparently a rat inside the machine, and while the machine was being cleaned, the rat, with its gasoline-soaked fur, ran out of the machine and onto the heater. The heater promptly lit the rat's fur on fire, and the rat then ran back inside the coin-operated machine. The machine, laced with plenty of gasoline, subsequently exploded.

It was argued that the risk of a flaming rat was not reasonably foreseeable (which indeed, it would not have been), and therefore a reasonable person would not have avoided it successfully. The court, however, still imposed liability because the \textit{kind} of accident that happened (gasoline causing an explosion) \textit{was} reasonably foreseeable and should have been prevented, even if the precise way that the explosion happened included some rather unforeseeable murine players. 

The fact that a foreseeable risk played out in a very unforeseeable way did not lessen or change the liability. In other words, the court ruled that no reasonable person (or in this case, company) would permit the use of large quantities of gasoline to clean machines. There was a very foreseeable risk of danger when gasoline was used in this way, and it does not matter that the flaming rat in itself was an unforeseeable vehicle through which the danger was realized in this particular instance. 

Can AI foresee risks reasonably, as we would expect a human to? Conversely, is its own behavior a foreseeable risk for its human handlers?

This is another important area where AI practitioners and researchers can meaningfully inform AI governance -- which risks should be foreseeable and which shouldn't, when contending with AI behavior? If AI models of certain varieties are known to behave in unexpected ways, then the particular instantiation of flaming rat disaster that eventually results should not matter; erratic behavior is a foreseeable risk, and therefore deploying the AI is a foreseeable risk that should be undertaken only with reasonable precautions, if at all.


\subsection{Consumer expectations and reasonable alternative designs}
Within the realm of tort liability lies product liability, where reasonableness comes up once again. As AI is increasingly incorporated into products and even marketed as a stand-alone product, product liability doctrine is an important place to look for clues regarding how its faults will be litigated.

One type of product flaw is a design defect. Courts recognize two main ways of determining if a product has a design defect: \\

\begin{enumerate}
    
    \item If the product fails to meet an ordinary consumer's expectations of safety (the consumer expectations test) \citep{kysar2003expectations} 
    \item If there was a reasonable alternative design that the makers failed to use instead (the reasonable alternative design test) \citep{westerbeke1998reasonable}. 
\end{enumerate}

These two tests call for reasonableness both in what a person can expect of an AI's behavior, and what AI creators can reasonably \textit{make} an AI do. Courts often incorporate both into their decision-making, and the consumer expectations test has maintained considerable staying power despite the more concrete implied risk-utility calculation for reasonable alternative design  \citep{kysar2003expectations}. 

There is perhaps good reason for this, and something we should note as AI practitioners: reasonableness, at the end of the day, should conform to what the people using (and interacting with) AI models expect of reasonable behavior and safety. 

The errors in an AI's behavior, if not in accordance with consumer expectations, violate this basic product safety test. This is where the \textit{nature} of the errors, not just the \textit{frequency} of the errors, becomes important in a way not accounted for in quantitative metrics for AI -- while we almost always quantitatively measure model accuracy and error rates, we do not often conduct in-depth behavioral analyses that characterize and study the nature of deviations from expected or reasonable behavior in practice. 

There is ample evidence that even the most advanced foundation models and large language models (LLMs) are misaligned with consumer expectations. This mismatch can be seen with even the most legally-knowledgeable consumers -- two lawyers were recently fined by a Manhattan judge for filing a legal brief that included made-up cases created by ChatGPT's hallucinations \cite{chatgpt-lawyers}. Clearly what consumers are expecting is not in line with what the models are producing, and there are serious potential consequences of such misalignment. The public is already using LLMs without understanding that they lie, hallucinate, and interpolate in ways a human collaborator would not. 

That said, when there are reasonable alternative designs available (for example, race- and gender-bias-audited algorithms \cite{ny-bias-hiring}), then even under the reasonable alternative design principle (the easier test of the two for most AI creators), there will \textit{still} be very little sympathy for AI creators who choose not to exercise even these preliminary cautions.




\section{Criminal Law}




\subsection{``Beyond a reasonable doubt'' and intentionally qualitative standards}
The idea of reasonableness shows up in one of the most commonly recognizable phrases in law: jury instructions in a criminal trial, which require the jury to convict if and only if guilt has been proven ``beyond a reasonable doubt.'' This is in contrast to civil cases, including the tort cases discussed above, which only require a preponderance of the evidence (more likely than not) standard.

Interestingly for us quantitative researchers, the law has explicitly declined to quantify what a ``reasonable'' doubt means, at least in attempts to crudely or reductively simplify the standard by using numbers. For example, in \citet{mccullough}, the trial judge included in his jury instructions ``a scale of zero to ten.'' He then calibrated by placing the preponderance of the evidence standard from civil trials at a bit over five out of ten, and instructed a jury that ``beyond a reasonable doubt'' was ``seven and a half, if you had to put it on a scale.'' The Nevada Supreme Court reversed, specifically restoring reasonableness to a standard that could not be simplistically quantified: 

\begin{quote}
    “The concept of reasonable doubt is inherently qualitative. Any attempt to quantify it may impermissibly lower the prosecution’s burden of proof, and is likely to confuse rather than clarify.” 
    
    \hspace{29mm}\parbox{\textwidth}{-\citet{mccullough}
    }
\end{quote}

\subsection{Statutory interpretation in criminal law}
Words are inherently ambiguous, and often the exact meaning of a word will be debated at length in court in a way that will decide the course of someone's life -- whether they are sent to prison, whether they lose their house, whether they can keep custody of their child, often rests on the court's interpretation of our collective understanding of a word in context. The stakes are perhaps highest in criminal law, where parties can disagree strongly on what a ``reasonable'' interpretation of a word may mean. 

In \citet{muscarello}, the matter in question was whether to ``carry'' a gun included carrying in your car and not on your person. The federal criminal code imposes a minimum 5-year mandatory prison sentence on anyone who ``uses or carries a firearm'' ``during and in relation to'' a ``drug trafficking crime.'' \cite{carry-gun-law}. The majority, who take a broad interpretation of ``carry,'' first cite the `ordinary English' definition, then the first definition in the Oxford English Dictionary, and then even examples of the word ``carry'' from the King James Bible. They even attempt to do some data analysis, saying ``to make certain that there is no special ordinary English restriction (unmentioned in dictionaries) upon the use of ``carry'' in respect to guns, we have surveyed modern press usage, albeit crudely, by searching computerized newspaper data bases.'' From their data analysis, they find that about one-third of the time, ``carry'' refers to carrying outside one's person. 

Justice Ginsburg, in her dissenting opinion, takes each of the majority's chosen sources and illustrates the prevalence of a much narrower definition of ``carry,'' to support her argument that the majority is cherrypicking examples. She then makes a compelling data science argument about reasonable behavior, pointing out the implication of the majority's own data analysis: if only one-third of the instances of the word support their argument, then presumably \textit{two}-thirds of common uses of ``carry'' do indeed refer to carrying only on one's person.

This is one prominent example of something that comes up incessantly in every aspect of the law, and perhaps most critically in criminal law -- the interpretation of words and phrases in laws and policies. As we work toward reasonable behavior for AI, we will have to define some tricky boundaries for what interpretation of a word, in context, is ``reasonable.'' Could an interpretation be reasonable even if only 1 out of 100 people would interpret it this way, and should they be punished for that interpretation? Do we apply a different, stricter standard to AI behavior? The most important lesson from such cases is the need for coherent debates regarding word meanings, and for this we need greater transparency for models' word meanings at the conceputal \cite{rane2024concept} and representational levels \cite{sucholutsky2023getting}. In Muscarello, while the justices disagree on the meaning of ``carry'' in context, they are able to use many other words to successfully \textit{disagree} about it, and that is the key to constructive human discussion and debate. In our discourses, we agree sufficiently on many more shared word meanings than the ones we disagree on, and therefore we can use words to constructively disagree about words, which in itself is an example of reasonableness. We must work towards being able to do the same with AI.

\subsection{Intent and Mens Rea}
In criminal law, a person's mental state is a crucial part of the assessment of whether they committed a crime and if so, how severe the penalty should be. \textit{Mens rea}, or mental state as referred to in criminal law, is subject to deep and thorough scrutiny along various axes: Did the person purposely and knowingly commit an act, or were they instead reckless or negligent? Were they exercising willful blindness, because they believed there was a high probability they could get in more trouble if they had full knowledge? If so, they are subject to more severe penalties than if they lacked knowledge for more innocuous, reasonable reasons \cite{criminal-law-textbook}. 

The problem with governing intent in AI behavior becomes immediately clear to us as AI practitioners: What does ``intent'' mean for a model? How can we define a model's mental state, even if its behavior is very human-like? If a human tries to absolve their own responsibility by relegating some controversial decision-making to an AI, should we evaluate the human's mental state or the model's, and what criteria can we use to evaluate either?



\subsection{Reasonable person standard for use of police force}

The courts have frequently enforced a reasonableness standard for reviewing police use of force \cite{terrill2009elusive, rice, graham}. In particular, courts have emphasized that police behavior should be evaluated ``from the perspective of a reasonable officer on the scene, rather than with the 20/20 vision of hindsight.'' \cite{graham}.  

With law enforcement increasingly relying on algorithmic decision-making, we are forced to ask how this reasonableness standard can possibly apply to evaluating commericalized predictive policing algorithms in which human beings are surveilled based on an opaque, proprietary point-based score for likelihood of committing a crime in the future \cite{predpol}. If qualified immunity for law enforcement officers also (often unbeknownst to the public) is implicitly extended to the outputs of the hidden algorithms they use to make decisions, the stakes for the behavior of those algorithms are higher than ever. 

Perhaps a better question is whether the reasonableness standard should apply not only to the \textit{behavior} of these algorithms, but also to their \textit{use}, in this case by law enforcement. Particularly in criminal law where the stakes for those involved are unimaginably high, the reasonableness standard for AI behavior should be only a \textit{lower} bound -- algorithms should be subject to, at the very \textit{least}, the same bounds on behavior as their human counterparts, and ideally should be subject to many more restrictions. Most importantly, for us all to have a democratic say in how we are policed, these algorithms and scores, if used at all, should be easily understandable to laypersons who are subject to their surveillance and arbitrary decision-making. This kind of transparency is the first step towards being able to challenge AI decision-making the same way we can at least challenge abuses of power (\textit{un}reasonable behavior) from human decision-makers today.

\section{Limitations and dangers of a reasonableness standard for AI}
The legal system's outcomes are obviously far from perfect -- one only has to look back a century or two to find a time when women being lawyers would be considered ``unreasonable'' -- and yet it is also a system in which, thankfully, our collective societal definitions of what is reasonable are allowed to evolve as we learn more about ourselves and about society. In this section we briefly discuss the limitations of the reasonableness standard as applied to AI behavior. 






\subsection{When we \textit{don't} want reasonableness}
There are cases where AI models have already shown tremendous success precisely because they provide solutions \textit{outside} the scope of solutions humans would ordinarily produce. There are clear reasons to allow and even encourage AI used for this kind of innovation \cite{alphago, alphafold}. As long as these models are contained to their specific use-cases, contained within the realms of a Go board and protein-folding simulation, and not let loose to interact with humans while trained to imitate a deceptively human-like manner, they are useful in their own right and generally not a threat.

However, when any such models come into contact with society at large, either as an LLM, a home robot, or a self-driving car, and \textit{especially} as they are increasingly powerful enough to cause real harm to the humans they interact with, they \textit{must} be held accountable. This includes models that may be simple in nature (or proprietary), but are nonetheless powerful in the human \textit{decisions} they are replacing or augmenting. We must hold them to (at least) the same standard we apply to humans in the same situations (the social worker deciding state aid \cite{socialworkerstateaid}, the nurse making a sepsis diagnosis \cite{nursesepsis}, the judge or jury determining sentencing \cite{compas}); these are the increasingly widespread cases in which the reasonable person standard can be useful. 



\subsection{Challenging reasonableness in AI decision-making}
Our legal system is adversarial by design; this is motivated by the idea that if two parties apply their best efforts in an adversarial process, the truth will be forced to emerge. While this system is far from equal in opportunity, its adversarial nature is intended to allow every person the opportunity to fight for their rights. Core to the idea of reasonableness in the legal system is the idea that reasonableness can and should be challenged and debated. 

Interpretability and explainability provide an important research agenda towards this goal, but are not nearly sufficient. In order for everyone to be able to challenge the notion of what ``reasonable'' behavior \textit{should} look like in AI systems, the interpretability and explainability of such systems should be easily accessible to an audience of laypersons. 

\subsection{Geographic and cultural implications}
Even the most well-intentioned legal principles or policy goals, when transplanted from one context to another, lead to unexpected and often dangerous outcomes. The notion of ``reasonable'' behavior is no different. For example, in AI for conservation efforts (extremely well-intentioned, and usually aimed towards social good), one use case involves AI models analyzing UAV footage from national parks in Africa to detect poachers. This might be a good start, but the local reality for many farms who share a border with protected wildlife refuge areas is that farmers often have to shoot predatory animals who attack their livestock along these borders. An AI system built in a lab in the U.S., according to the standards of reasonableness we apply here, is unlikely to automatically consider the difference between a poacher and a farmer protecting his livestock (and indeed it is a good idea to ask whether the AI should be the entity making such a judgement call in the first place). As models deployed globally continue to be built in a few concentrated geographic locations, these mismatches are going to become increasingly burdensome on local communities who often have no say in model development. Incorporating the voices of local communities early on, and keeping the models grounded to the realities in which they operate, will be incredibly important. 

\section{Cognitive science and social science to provide an understanding of reasonable human behavior}

Cognitive science and social science provide us with empirical analyses of human behavior that will be critical to defining and refining what the distribution of ``reasonable'' behavior looks like in various contexts. Importantly, while it is absolutely critical to take a data-driven approach towards refining the reasonable person standard for AI, this is not something that can be neatly quantified for all contexts ex ante. Rather, as in many other fields, it calls for what \citet{newell2007computer} call ``laws of qualitative structure.'' Reasonable behavior, and indeed most of human behavior, is not easily quantified in the traditional sense \cite{simon1990invariants} -- certainly not in the way we have done in machine learning in recent years, with a small set of quantitative metrics run on a large swath of models. Instead, a much deeper dive is needed into each model type so that we may scratch the surface below it's initial behavior and truly characterize it with respect to the reasonable person standard for humans. 

Characterizing the full scope, distribution, and diversity of human-like behavior is not easy; it is an ongoing practice even for humans (the entire fields of cognitive science, psychology, and neuroscience are devoted to this study). And yet it is an increasingly, urgently necessary line of inquiry for AI behavior. To begin meaningful inquiry, we don't have to aim for a perfect solution; even small steps towards a qualitative, reasonableness-based assessment of AI, empirically grounded in data from both humans and models, is a good place to start. 


\section*{Acknowledgements}
I am deeply grateful to my many colleagues and mentors, in law, cognitive science, and computer science, who have been so generous with their time and feedback in shaping this paper. 
\nocite{langley00}

\bibliography{example_paper}
\bibliographystyle{icml2024}




\end{document}